\documentclass[12pt]{iopart}

\usepackage{iopams}  
\usepackage{graphicx}

\begin{document}

\title{Effect of n- and p-type Doping on Coherent Phonons in GaN}

\author{Kunie Ishioka$^1$, Keiko Kato$^1$ \footnote{Present address: NTT Basic Research Laboratories}, Naoki Ohashi$^1$, Hajime Haneda$^1$, Masahiro Kitajima$^1$ \footnote{Present address:National Defense Academy} and Hrvoje Petek$^2$}

\address{$^1$ National Institute for Materials Science, Tsukuba, 305-0047 Japan}
\address{$^2$ Department of Physics and Astronomy, University of Pittsburgh, Pittsburgh, PA 15260, USA}
\ead{ishioka.kunie@nims.go.jp}

\begin{abstract}

Effect of doping on the carrier-phonon interaction in wurtzite GaN is investigated by pump-probe reflectivity measurements using 3.1 eV light in near resonance with the fundamental band gap of 3.39 eV.  Coherent modulations of the reflectivity due to the $E_2$ and the $A_1$(LO) modes, as well as the 2$A_1$(LO) overtone are observed.  Doping of acceptor and more so for donor atoms enhances the dephasing of the polar $A_1$(LO) phonon via coupling with plasmons, with the effect of donors being stronger.  Doping also enhances the relative amplitude of the coherent $A_1$(LO) phonon with respect to that of the high-frequency $E_2$ phonon, though it does not affect the relative intensity in Raman spectroscopic measurements.  This enhanced coherent amplitude indicates that transient depletion field screening (TDFS), in addition to impulsive stimulated Raman scattering (ISRS), contribute to generation of the coherent polar phonons even for sub-band gap excitation.  Because the TDFS mechanism requires photoexcitation of carriers, we argue that the interband transition is made possible at the surface with photon energies below the bulk band gap through the Franz-Keldysh effect.

\end{abstract}

\maketitle

\section{Introduction}

\begin{figure}
\includegraphics[width=0.5\textwidth]{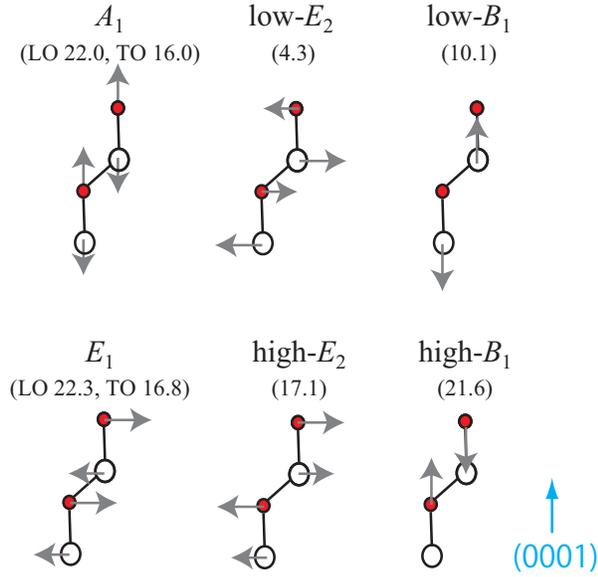}
\caption{\label{phononmode} 
(Color Online.) Atomic displacements of the $\Gamma$-point phonons in wurtzite GaN \cite{Zhang97}.  Gallium and nitrogen atoms are colored white and red.  Low-$E_{2}$ (low-$B_{1}$) and high-$E_{2}$ (high-$B_{1}$) denote the low- and high-frequency $E_2$ ($B_1$) modes, respectively.  
Numbers in parentheses denote the frequencies in THz obtained in previous studies \cite{Azuhata95,Zhang97}.
}
\end{figure}

Gallium nitride (GaN) is an important optoelectronic semiconductor with a direct-gap of 3.39 eV at room temperature.  The optical properties of GaN depend on the rapid exchange of energy between the electronic and the lattice subsystems \cite{Kasic00,Song06,Tsen06}.  The wurtzite polymorph of GaN has optical phonon modes of $A_1$ (Raman (R) and infrared (IR) active), $E_1$ (R and IR active), $E_2$ (R active) and $B_1$ (silent) symmetries \cite{Kozawa94,Azuhata95,Zhang97}, whose atomic displacements are summarized in Fig. \ref{phononmode}.    The frequencies of the IR active $A_1$ and $E_1$ modes split into a longitudinal optical (LO) and a transverse optical (TO) components by the macroscopic electric field associated with the LO phonon.  This electric field stiffens the force constant and thereby increases the LO frequency over the TO.  For wurzite GaN the LO-TO splitting is greater than the $A_1-E_1$ splitting, $i.e., |\Omega_{LO}^{(A1)}-\Omega_{TO}^{(A1)}|$ and $|\Omega_{LO}^{(E1)}-\Omega_{TO}^{(E1)}| \gg |\Omega_{LO}^{(A1)}-\Omega_{LO}^{(E1)}|$ and $|\Omega_{TO}^{(A1)}-\Omega_{TO}^{(E1)}|$,  indicating that the electrostatic force dominates over interatomic anisotropy in the short-range forces \cite{Arguello69}.  

Like in zinc-blende crystals, the LO phonons in wurtzite GaN can couple with plasmons to form LO phonon-plasmon coupled (LOPC) mode \cite{Kozawa94,Perlin95,Harima98,Wieser98,Demangeot98}.  The $A_1$- and $E_1$-like LOPC modes can be distinguished by the detection geometry and optical polarization in Raman scattering measurements \cite{Wieser98}.  
In general, the frequencies of the LOPC mode are given by the solution to the equation for the frequency-dependent dielectric response of the lattice and the free electrons \cite{Fukasawa94,Irmer97}:
\begin{equation}
\varepsilon(\omega)=\varepsilon_{\infty}\Bigl[1+\frac{\Omega_{LO}^2-\Omega_{TO}^2}{\Omega_{TO}^2-i\Gamma\omega-\omega^2}-\frac{\omega_p^2}{\omega^2+i\gamma\omega}\Bigr]=0,
\label{plasma1}
\end{equation}
where $\varepsilon_\infty$ is the high frequency dielectric constant, $\gamma$ and $\Gamma$, plasmon and phonon damping rates, and $\Omega_{LO}$ and $\Omega_{TO}$, the LO and TO phonon frequencies.   The plasma frequency,
\begin{equation}
\omega_p=\sqrt{\frac{ne^2}{m^*\varepsilon_0\varepsilon_\infty}},
 \label{plasma2}
\end{equation}
depends on the free carrier density, $n$ their effective mass, $m^*$, and the vacuum dielectric constant, $\varepsilon_0$.  For n-type GaN, the LOPC mode frequencies measured by Raman spectroscopy were found to follow the well-known undamped ($\gamma=\Gamma$=0) solutions of eq. (\ref{plasma1}) \cite{Kozawa94,Harima98,Perlin95,Wieser98}:
\begin{equation}
2\omega_{\pm}^2=\omega_p^2+\Omega_{LO}^2\pm[(\omega_p^2+\Omega_{LO}^2)^2-4\omega_p^2\Omega_{TO}^2]^{1/2}.
\label{plasma}
\end{equation}
For p-type GaN, by contrast, the coupling with plasmons was observed only as a slight broadening of the Raman LO peak without significant frequency shift \cite{Harima98}.  The difference between the n- and p-type GaN was attributed to the heavy damping ($\gamma\gg\omega_p$) of the hole plasmon.

Detection of coherent optical phonons through transient transmission or reflectivity measurements enables monitoring the femtosecond time evolution of carrier-phonon coupling.  Depending on the excitation photon energy, the excitation pulses may or may not add photocarriers to the valence and conduction bands. A previous transient transmission study from GaN(0001) surface with nonresonant 1.5 eV light \cite{Yee02} observed the two $E_2$-symmetry modes and the $A_1$(LO) mode.  Only the polar $A_1$(LO) mode exhibited a pump power-dependent dephasing rate, confirming its strong Fr{\"o}hlich interaction with the carriers photoexcited via a nonlinear three-photon process.  No coherent LOPC mode, whether coupled with chemically doped or photoexcited carriers, was reported. 

In the present study, we examine the electron-phonon coupling in differently doped GaN single crystals by transient reflectivity measurements with 3.1 eV light.  With sub-band gap excitation we expect the coherent phonons to be excited mainly via nonresonant Raman process; the excitation of electron-hole pairs below the band gap resonance could be made possible only by Franz-Keldysh effect within the surface depletion region.  We find that the dephasing rate of the $A_1$(LO) phonon is increased and its frequency upshifted by doping with donor atoms, whereas the dephasing rate is increased without any frequency shift by doping of acceptor atoms.  We attribute our observation to the formation of the LOPC mode coupled with chemically doped carriers.  We also observe the coherent amplitude of the $A_1$(LO) phonon to be enhanced significantly by the doping.  Comparison with Raman spectra of the same samples suggests photoexcited ultrafast current in the surface depletion region is responsible for the enhancement.  

\section{Experimental}

The samples studied are GaN single crystal wurtzite polymorphs grown by metal organic chemical vapor deposition (MOCVD) on (0001)-oriented sapphire substrates \cite{Yao10}.  The nominally undoped GaN layer is 50 $\mu$m thick and has a defect concentration of $\sim1\times10^{16}$cm$^{-3}$, which makes it slightly n-type.  The Si-doped n-type GaN film of 3.4 $\mu$m thickness is grown with Si concentration of $\sim1\times10^{18}$ cm$^{-3}$ on a 1.6 $\mu$m thick undoped buffer layer.  The Mg-doped p-type GaN film of 0.5  $\mu$m thickness is grown with Mg concentration of $\sim1\times10^{18}$cm$^{-3}$ on a 5 $\mu$m thick undoped layer.  For comparison we also investigate an (0001)-oriented semi-insulating (SI) GaN single crystal of 1mm thickness (Kyma Technologies Inc.).  

Pump-probe reflectivity measurements are performed with optical pulses with $\sim$10 fs duration, 400 nm wavelength (3.1 eV energy), and 70 MHz repetition rate.  Only a very small fraction ($\sim0.1\%$) of the optical pulse spectrum exceeds the fundamental gap of GaN.  
Linearly polarized pump and probe beams are incident with angles of 20$^\circ$ and 5$^\circ$ from the surface normal.  In this near back-reflection geometry from the (0001) surface, only the $E_2$ and $A_1$(LO) modes are dipole-allowed by the Raman scattering selection rules, which also dominate the generation and detection of coherent phonons \cite{Ishioka11}.  The pump-induced change in the reflectivity $\Delta R$ is measured by detecting the probe beam reflected from the sample surface (isotropic detection) and accumulating the signal in a digital oscilloscope while scanning the time delay between the pump and the probe pulses at 20 Hz (fast scan).   

\begin{figure}
\includegraphics[width=0.65\textwidth]{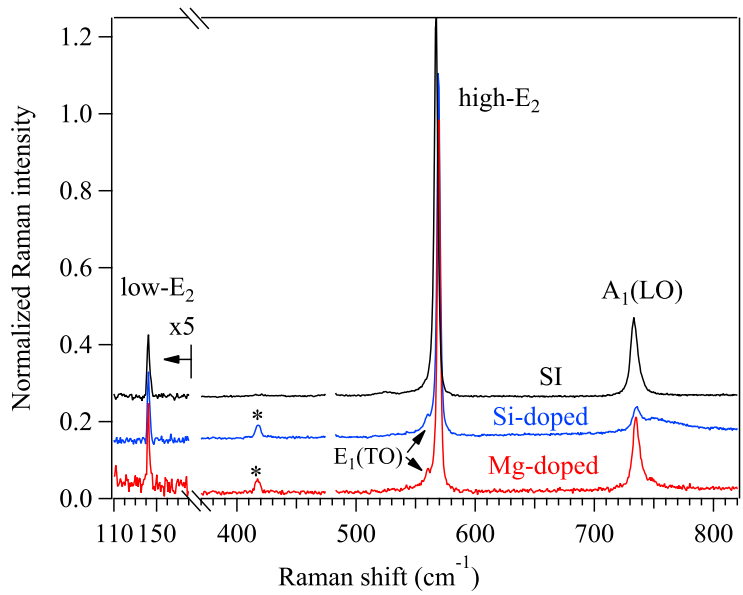}
\caption{\label{Raman} 
(Color Online.) Raman spectra of the differently doped GaN samples.  The Raman intensities are normalized by the height of the high-$E_2$ phonon peak.  Traces are offset for clarity.  Asterisk (*) indicates the $A_{1g}$ phonon of the sapphire substrate.
}
\end{figure}

Separate Raman scattering measurements on the same samples are performed in back-reflection geometry using a Raman microscope with nonresonant 532-nm (2.3-eV) excitation.  The low-$E_2$, high-$E_2$ and $A_1$(LO) phonons are observed at 142, 569 and 736 cm$^{-1}$, as shown in Fig. \ref{Raman}; these comprise all the fundamental phonon modes allowed in the back-scattering geometry from the (0001) surface.  For the Si- and Mg-doped samples the $A_1$(LO) peak splits into two, $i.e.$, the $A_1$(LO) phonon coupled with plasmons ($A_1$-like LOPC mode) from the doped layer and the bare $A_1$(LO) phonon from the undoped buffer layer \cite{Kozawa94,Perlin95,Harima98,Wieser98,Demangeot98}.  For the thin film samples small Raman peaks from the dipole-forbidden $E_1$(TO) phonon of GaN and from the sapphire substrate are also observed at 561 and 417 cm$^{-1}$.  
	
\section{Results and discussion}

\begin{figure}
\includegraphics[width=0.65\textwidth]{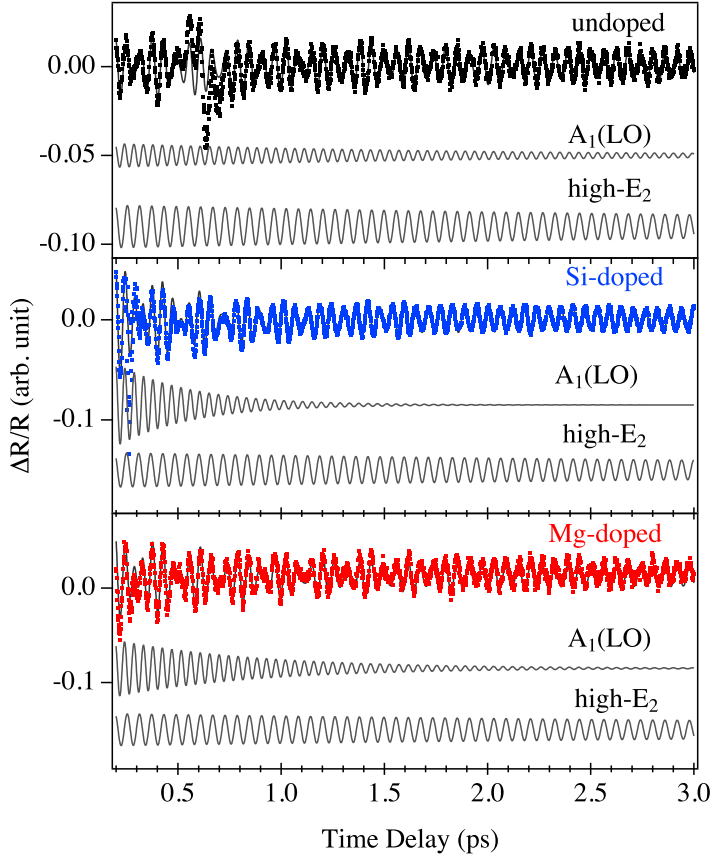}
\caption{\label{TD} 
(Color Online.) Oscillatory parts of transient reflectivity responses $\Delta R/R$ (dots) and their decomposition into components due to the coherent $A_1$(LO) and high-$E_2$ modes (solid curves) for differently doped GaN samples.
}
\end{figure}

\begin{figure}
\includegraphics[width=0.75\textwidth]{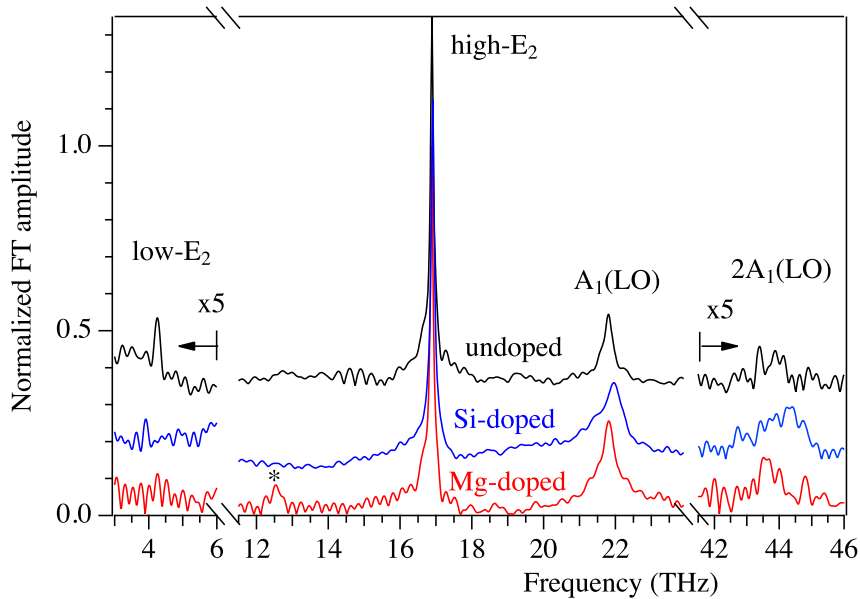}
\caption{\label{FT} 
(Color Online.) Fourier transformed (FT) spectra of the oscillatory reflectivity change of differently doped GaN samples.  The FT amplitude is normalized by the height of the high-$E_2$ phonon peak.  Traces are offset for clarity.  Asterisk (*) indicates the $A_{1g}$ phonon of the sapphire substrate.
}
\end{figure}

Figure \ref{TD} compares the pump-induced reflectivity change $\Delta R/R$ of differently doped GaN samples.  After photoexcitation at $t=0$, the reflectivity oscillates with a clear beating pattern.  The beating is mainly contributed by the high-frequency $E_2$ mode at 17 THz and the $A_1$(LO) mode at 22 THz, as shown in the Fourier transform (FT) spectra in Fig. \ref{FT}.  The individual oscillation components extracted by fitting the time-domain data to a sum of damped harmonic oscillations:
\begin{eqnarray}
\frac{\Delta R(t)}{R}&\simeq\frac{\Delta R_A}{R}\exp(-\Gamma_A t)\sin(2\pi\Omega_A t+\psi_A)\nonumber\\
&+\frac{\Delta R_E}{R}\exp(-\Gamma_E t)\sin(2\pi\Omega_E t+\psi_E),
\label{ddh}
\end{eqnarray}
are also shown in Fig. \ref{TD}.  The observation of the $E_2$ and $A_1$-symmetry modes in the near back-reflection configuration is expected from the Raman selection rules \cite{Yee02,Azuhata95}. 
We also observe a weak low-$E_2$ mode at 4 THz and an overtone of the $A_1$(LO) phonon (2$A_1$(LO)) at 44 THz, as shown in Fig. \ref{FT}.  The small amplitude of the 2$A_1$(LO) mode is in contrast to the resonant Raman spectrum at wavelengths shorter than 380 nm, which was found to be dominated by the overtones up to 7th \cite{Wieser98,Kaschner01,Sun02}.  Because our measurement system has sufficient bandwidth to detect the coherent response with bandwidth exceeding 100 THz \cite{Hase12}, the weak overtone signal can be attributed to a pre-resonant response to photoexcitation at 400 nm.   our observation thus indicates that the resonant effect is negligible with photoexcitation at 400 nm.  The origin of the 2$A_1$(LO) mode must be different from the recently reported harmonic frequency comb generation in Si through the amplitude and phase modulation of the reflected light within the optical skin depth, because the GaN samples are transparent to 3.1 eV light \cite{Hase12}. With increasing pump power, the amplitudes of all the coherent phonon modes increase linearly, while their dephasing rates and the frequencies exhibit no systematic pump-power dependence for all the GaN samples examined in the present study.  This confirms that multi-photon excitation of carriers is negligible under our photoexcitation with 3.1 eV light, in contrast to the near infrared photoexcitation in the previous transmission study \cite{Yee02}.  

\begin{figure}
\includegraphics[width=0.55\textwidth]{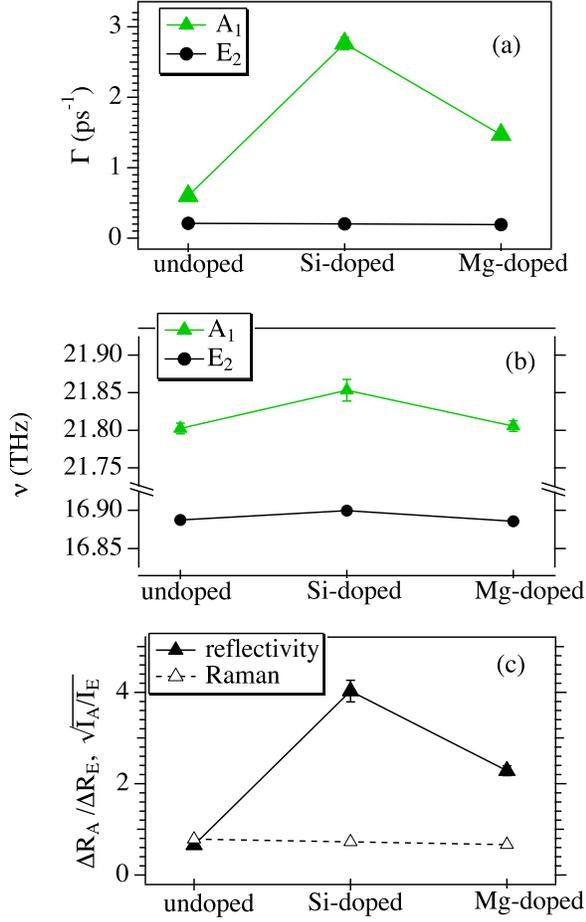}
\caption{\label{GammaR} 
(Color Online.)  (a) Dephasing rates and (b) frequencies of coherent high-$E_2$ and $A_1$(LO) phonons for differently doped GaN samples obtained from transient reflectivity measurements.  (c) Relative coherent phonon amplitudes $\Delta R_{A}/\Delta R_{E}$ and square root of the Raman intensity ratios $\sqrt{I_{A}/I_{E}}$ for differently doped GaN samples.  The Raman intensity $I_{A}$ is the sum of the bare $A_1$(LO) and the LOPC modes.
}
\end{figure}

Doping of GaN by Si and Mg impurities modifies the coherent $A_1$(LO) mode significantly　through the coupling with plasmons, whereas the high-$E_2$ mode is hardly affected, 
as summarized in Fig. \ref{GammaR}.  The dephasing rate of the $A_1$(LO) mode is $\Gamma_A\sim$0.6 ps$^{-1}$ for the undoped sample, which is consistent 
with the lifetime obtained by time-resolved Raman spectroscopy at a low photocarrier density (10$^{16}$ cm$^{-3}$) \cite{Tsen06}.  Doping of Si impurities (n-type) increases $\Gamma_A$ to 2.8 ps$^{-1}$ and upshifts the frequency $\Omega_A$ by 0.05 THz 
with respect to that of the undoped sample.  Doping of Mg (p-type) also enhances $\Gamma_A$ but the effect is smaller than n-doping, and $\Omega_A$ is hardly affected, as shown in Fig. \ref{GammaR}(a,b).  Our observations are consistent with the previous Raman studies, which concluded that the coupling of the LO phonon with hole plasma is less efficient than that with electron plasma due to the stronger damping of holes \cite{Harima98,Demangeot98}.  In general, the LO phonons can couple with both chemically doped and photodoped electrons and holes.  In the present study, however, we consider that the LO phonons couple mainly with chemically doped electrons and holes for Si- and Mg-doped samples, because we observe comparable line width for the A1-like LOPC peak in our Raman scattering spectra [Fig. \ref{Raman}]. We note that in time domain we observe only the coherent $A_1$-like LOPC mode for the n- and p-doped layers [Figs. \ref{TD},\ref{FT}], though in the Raman spectrum we clearly see the bare A1(LO) phonon from the undoped buffer layer as well [Fig. \ref{Raman}].  This is probably because the coherent $A_1$(LO) phonons are generated much more efficiently for doped layers than for undoped one, as we will discuss in the following paragraph.
	
\begin{figure}
\includegraphics[width=0.6\textwidth]{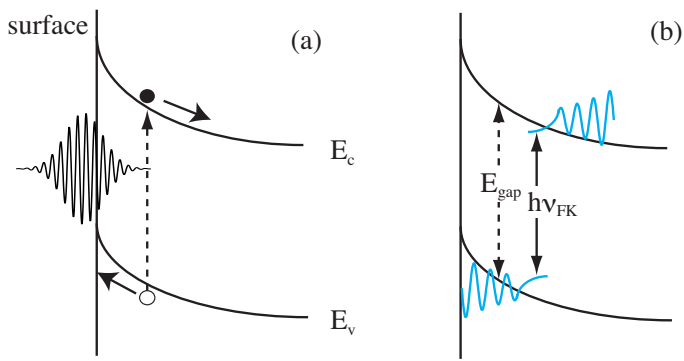}
\caption{\label{TDFS} 
(Color Online.) Schematic illustrations of the band structure of an n-type semiconductor surface.  (a) Photoinjected current screens the built-in surface field and excites coherent polar phonons via TDFS \cite{Foerst}.  (b) Franz-Keldysh effect involving the coupling between the tails of the electron and hole wavefunctions  causes the interband transition at an energy $h\nu_{FK}$ smaller than the band gap $E_{gap}$ \cite{Cavallini07}.
}
\end{figure}

Doping of Si or Mg also significantly enhances the amplitude of the $A_1$(LO) mode relative to that of the high-$E_2$ mode, i.e., $\Delta R_{A}/\Delta R_{E}$, whereas the Raman peak intensity ratio $I_A/I_E$ is unchanged, as shown in Fig. \ref{GammaR}(c).  The difference can be explained in terms of the generation mechanism of the coherent $A_1$(LO) phonon.  If all the coherent phonons are solely generated  via impulsive stimulated Raman scattering (ISRS) \cite{Dhar}, the relative amplitudes should be proportional to the square root of the relative Raman intensity $\sqrt{I_{A}/I_{E}}$ measured at the same wavelength.  In resonant photoexcitation on polar semiconductors, however, coherent polar optical phonons can also be generated via transient depletion field screening (TDFS) mechanism \cite{Foerst}, in which ultrafast screening by photoinjected current induces an abrupt change in the surface built-in field [Fig. \ref{TDFS}(a)].  In the present study, photocarriers cannot be generated in the bulk GaN, since the photon energy is slightly below the fundamental band gap.  Nevertheless, in the surface depletion region of n-doped GaN, the steep band bending can induce tails of the valence and conduction bands of about 0.1 eV through the Franz-Keldysh effect \cite{Cavallini07}, as schematically illustrated in Fig. \ref{TDFS}(b).  This would allow a small but finite fraction of the 3.1 eV light to excite the interband transition.  The photoexcited electrons would drift deeper into the bulk region, and thereby generate coherent $A_1$(LO) phonons by TDFS  and form the LOPC mode.  We thus attribute the enhancement in the coherent LOPC amplitude by n-doping to the additional TDFS generation.  Similar mechanism can be at work at the surface of a p-type semiconductor.  The difference in the generation efficiency for the n- and p-doped samples depends on the position of the Fermi level pinning with respect to the band gap edges.  
	
\section{CONCLUSION}

We have investigated the effect of n- and p-type doping on carrier-phonon dynamics of GaN under photoexcitation with 3.1 eV light.  The dephasing rate, the frequency and the amplitude of the polar $A_1$(LO) phonon are modified by coupling with chemically doped electron (hole) plasma for n-doped (p-doped) sample, with the effect of n-doping greater than that of p-doping.  Because the photon energy is slightly short of the band gap of bulk GaN, photoexcitation is made possible by Franz-Keldysh effect only at the surfaces with steep band bending.  Such photoexcitation enables the generation of coherent $A_1$(LO) phonons via transient screening in the surface depletion field.  Our study reveals the complex interactions between the carriers and lattice affecting the optical properties of GaN in the surface band bending region.

\ack{This research was partially supported by  NSF CHE-1213189 grant.}
\\
\bibliographystyle{unsrt}
\bibliography{GaN}

\end{document}